\documentclass[aps,twocolumn,pra]{revtex4-1}
\usepackage{graphics}
\usepackage{dcolumn}
\usepackage{bm}
\usepackage{amsmath}
\usepackage{hyperref}
\hypersetup{colorlinks=true, urlcolor=blue}
\usepackage{color}
\usepackage{tikz}
\usepackage{multirow}

\usepackage{float}
\usepackage[caption = false]{subfig}
\usepackage{newlfont}
\usepackage{amssymb}
\usepackage{amsfonts}
\usepackage{amsthm}
\usepackage{graphicx}
\usepackage{bm}

\def \be{\begin{equation}}
\def \ee{ \end{equation} }

\definecolor{myurlcolor}{rgb}{0,0,0.7}
\hypersetup{ colorlinks,linkcolor=myurlcolor,citecolor=myurlcolor,urlcolor=myurlcolor}

\begin{document}
\title{Family of Quantum Mutual Information in Multiparty Quantum Systems}

\author{Asutosh Kumar}
\email{asutoshk.phys@gmail.com}

\affiliation{Department of Physics, Gaya College, Rampur, Gaya 823001, India}
\affiliation{P.G. Department of Physics, Magadh University, Bodh Gaya 824234, India}

\begin{abstract}
The characterization of information within a multiparty system is both significant and complex. This paper presents the concept of generalized conditional mutual information, along with a family of multiparty quantum mutual information measures. We provide interpretations and delineate the properties of these concepts, while also pointing out certain unresolved issues. The generalized conditional mutual information serves to encapsulate the interdependencies and correlations among various components of a multiparty quantum system. Additionally, various formulations of multiparty quantum mutual information contribute to a deeper comprehension of classical, quantum, and total correlations. These insights have the potential to propel fundamental research in the field of quantum information theory.
\end{abstract}

\maketitle

\section{Introduction} 

In our information-driven universe, information is the cornerstone of communication, computation, and knowledge. Shannon in 1948 introduced the concept of {\em mutual information} which provided a framework for quantifying the amount of information shared between two systems, revolutionizing classical information theory. Classical mutual information $I(X:Y)=H(X) + H(Y) - H(X,Y)$, where $H(X)$ is the Shannon entropy, 
measures the amount of information obtained about one random variable through another random variable. It quantifies the reduction in uncertainty about one variable given knowledge of the other. In classical information theory \cite{thomas}, mutual information has diverse applications in data compression, error correction, and channel capacity, making it a fundamental tool in communication systems and coding theory.

This concept has now transcended into the quantum domain, playing a crucial role in quantum information theory \cite{nielsen, wilde-qit, preskill} in particular. Quantum mutual information (QMI) is a generalization of classical mutual information to quantum systems. The QMI of a bipartite quantum state $\rho_{AB}$ given by $I(A:B)=S(\rho_A) + S(\rho_B) - S(\rho_{AB})$, where $S(\rho)$ is the von Neumann entropy, 
quantifies the total correlation \cite{groisman, luo2007}, both classical and quantum, between subsystems $A$ and $B$. It serves as a measure of correlation beyond entanglement. 
QMI is essential in multiple areas of quantum information processing like quantum communication, quantum computing, and quantum cryptography \cite{nielsen, wilde-qit, preskill}. It is especially important in quantifying quantum channel capacities \cite{bennett, holevo}. In quantum machine learning \cite{biamonte, carleo}, it quantifies the information exchanged between various representations of quantum datasets. It is also significant as a probe for many-body localization \cite{tomasi}, and in quantifying quantum objectivity \cite{chisholm}. 

Understanding various correlations \cite{bengtsson, HHHH2009, modi2012, bera2018} in systems involving more than two parties becomes increasingly important. In classical information theory, mutual information between multiple random variables can be straightforwardly extended. In the quantum realm, however, correlations are more complex due to the exotic quantum phenomena such as nonlocality, superposition and measurement problem. Multiparty quantum mutual information (MQMI) extends the concept of bipartite QMI to systems with multiple parties, offering insights into the intricate correlations that arise in quantum states. It seeks to quantify the total correlations among several subsystems within a quantum state, and has implications for understanding and leveraging the correlations in multiparty quantum systems.
Mutual information and related measures \cite{watanabe, han1975, han1978, groisman, luo2007, walczak, giorgi2011, asu-qmi, sazim} are famed measures of multipartite information and correlation. 
The journey of mutual information from classical to quantum domains underscores its profound elegance and significance. As quantum technologies continue to advance, the significance of mutual information in both the realms becomes increasingly evident.
The two main contributions of this paper are: \\
1. Concept of {\em generalized conditional mutual information} which is an extension of the conditional entropy $S(A|B) = S(\rho_{AB}) - S(\rho_B)$ and conditional mutual information $I(A : B |C) = S(\rho_{AC}) + S(\rho_{BC}) - S(\rho_C) - S(\rho_{ABC})$. This quantity can encapsulate every possible interdependency and correlation of any subsystem of a multiparty system. \\
2. Introduction of a family of multiparty quantum mutual information. There are at least $n-1$ MQMI for an $n ~ (\ge 2)$-party quantum system. Moreover, any positive linear combination of these MQMIs is another MQMI. Some linear combinations of these MQMIs can yield negative values. \\
\indent
It is evident that multiple expressions of MQMI arise due to different ways of defining and quantifying correlations in complex quantum systems. Each expression may capture unique aspects of these correlations, leading to diverse applications and insights. 
A straightforward consequence of multiple expressions of MQMI would be enhanced understanding of classical, quantum and total correlations.
By providing a multifaceted understanding of correlations in multiparty quantum systems,
multiple expressions of MQMI can drive fundamental research in quantum information theory. By providing different perspectives on correlations and their operational interpretations, we can explore new theoretical models and deepen our understanding of quantum systems.

This paper is organised as follows. In Sec. \ref{sec:setup}, we consider the preliminaries such as notation and definitions. The notion of {\em generalized conditional mutual information} which is the generalization of conditional entropy and conditional mutual information to multiparty systems is introduced in Sec. \ref{sec:gcmi}. In Sec. \ref{sec:mqmi}, we introduce a family of multiparty quantum mutual information and provide their interpretations and properties. We also speculate them to be {\em secrecy monotones} which are useful in cryptography.
Finally, we conclude and discuss some unresolved issues in Sec. \ref{sec:conclusion}.

\section{Setup}
\label{sec:setup}
\subsection{Preliminaries}
We consider a multiparty quantum system $X_1X_2 \cdots X_n$ represented by finite dimensional density matrix  $\rho_{X_1X_2 \cdots X_n}$ $\in \mathcal{H}_1^{d_1} \otimes \mathcal{H}_2^{d_2} \otimes \cdots \otimes \mathcal{H}_n^{d_n}$. The reduced density matrix of subsystem $X$ is obtained by partial tracing over the remaining subsystems $\overline{X}$: $\rho_X = \text{tr}_{\overline{X}} (\rho_{X \overline{X}})$. 
The von Neumann entropy (in bits) given by
\begin{equation}
S(\rho) := - \text{tr} (\rho \log_2 \rho) = -\sum_i \lambda_i \log_2 \lambda_i,
\end{equation}
where $\lambda_i \ge 0$ and $ \sum_i \lambda_i = 1$, is the quantum counterpart of Shannon entropy. Shannon entropy is the average information content of a probability distribution. 
The von Neumann entropy satisfies the inequalities \cite{wehrl1978, ruskai}: $S(\rho_X) - S(\rho_Y) \le S(\rho_{XY}) \le S(\rho_X) + S(\rho_Y) \le S(\rho_{XZ}) + S(\rho_{YZ})$ and $S(\rho_{Y}) + S(\rho_{XYZ}) \le S(\rho_{XY}) + S(\rho_{YZ})$.
We denote the von Neumann entropy of subsystem $X_i X_j$ as $S(\rho_{X_i X_j}) \equiv S(X_i X_j) \equiv S_{ij}$, and so on. $S(\rho) = 0$ for a pure quantum state. 
Another important entropy for our purpose is quantum relative entropy (QRE) which measures the {\em closeness} of two density matrices. It is defined as
\begin{equation}
D(\tau || \sigma) := tr (\tau \log_2 \tau) - tr (\tau \log_2 \sigma),
\end{equation}
if $supp(\tau) \subseteq supp(\sigma)$, and infinity otherwise. 
The {\em support} of a Hermitian matrix is the Hilbert space spanned by its eigenvectors with nonzero eigenvalues \cite{nielsen}. The QRE is monotonic under partial trace, completely-positive and trace-preserving (CPTP) maps, and positive maps \cite{ruskai, hermes}.
Let $\mathcal{X} = \{X_1,X_2, \cdots ,X_n \}$, $[n] = \{1, 2, \cdots , n\}$, and $\mathbf{1} = 1$ such that $S(\mathbf{1}) = 0$ and for any system $X$, $\rho_X \mathbf{1} = \rho_X = \mathbf{1} \rho_X$. [One can alternatively consider $\mathbf{1} = \text{diag} \{1, 1, \cdots , 1\}$ with the requirement that its dimension is self-adjusting!]

\subsection{Correlations and Monotones}
Let us consider a function $f(\rho_{X_1 \cdots X_n})$ defined on $\rho_{X_1 \cdots X_n}$ and mention below some plausible and useful properties. \\ 
(P1) Symmetry: $f(\rho_{X_1 \cdots X_n})$ is symmetric under the interchange of any two parties $X_j$ and $X_k$. \\
(P2) Semipositivity: $f(\rho_{X_1 \cdots X_n}) \ge 0$. \\
(P3) Vanishing on product states: 
$f(\rho_{X_1} \otimes \cdots \otimes \rho_{X_n}) = 0$. \\
(P4) Monotonicity under some local operations [local (completely) positive maps]. \\
(P5) Monotonicity under classical communications (public announcement or communication over phone). \\
(P6) Additivity: $f(\rho \otimes \sigma) = f(\rho) + f(\sigma)$ and $f \left(\rho^{\otimes n} \right) = n f(\rho)$. \\
(P7) Continuity: $f(\rho)$ is a continuous (smooth) function of its argument $\rho$. \\
If $f$ satisfies (P2) semipositivity and (P3) vanishing on product density matrices, it is a measure of the amount of {\it correlation} between the parties. A nonnegative correlation function that observes (P4) monotonicity under local operations and (P5) monotonicity under classical communications is called a {\it monotone}. A {\it secrecy monotone}, in addition to (P2--P5), satisfies (P6) additivity and (P7) continuity.
If the secrecy monotone is expected to measure the amount of information (secrecy) shared by the communicating parties $\{X_1,X_2, \cdots ,X_n \}$ with the hostile party Eve, then the following properties are natural \cite{cerf, HHH2000}: \\
(P8) Monotonicity under local operations by Eve. \\
(P9) Monotonicity under classical (public) communication by Eve. \\
Whether the information in question is correlation, monotone or secrecy monotone should be clear from its properties and the context.

\section{Generalized Conditional Mutual Information}
\label{sec:gcmi}
In this section, we introduce the notion of generalized conditional mutual information (GCMI). It is the multiparty extension of conditional entropy $S(A|B) = S(AB) - S(B)$ and conditional mutual information $I(A : B |C) = S(AC) + S(BC) - S(C) - S(ABC)$. It can encompass every possible interdependency or correlation (interaction information) of any subsystem of a multiparty system.
We define the GCMI as the information contained in subsystems $X_{k_1}X_{k_2} \cdots X_{k_m}$ of a multiparty system 
$X_1X_2 \cdots X_n~~(n \ge m)$ but not in $Y$, where $Y$ (acting as a single system) is either $\mathbf{1}$ or one or more remaining subsystems,

\begin{align}
& I(X_{k_1}: X_{k_2}: \cdots :X_{k_m}|Y) := - S(Y)  \nonumber \\
&+ \sum^m_{j=1} (-1)^{j+1} \sum_{k_1 < \cdots <k_j \in [m]} S(X_{k_1} X_{k_2} \cdots X_{k_j}Y). \label{qci1}
\end{align} 
A few remarkable points about $I(X_{k_1}: X_{k_2}: \cdots :X_{k_m}|Y)$ are as follows. \\
1. $I(X_{k_1}: X_{k_2}|Y) \ge 0$. This follows from the strong subadditivity of von Neumann entropy.\\
2. $I(X_{k_1}X_{k_2} \cdots X_{k_m}|\mathbf{1}) = S(X_{k_1}X_{k_2} \cdots X_{k_m})$. \\
3. $I(X_{1}: X_{2}: \cdots :X_{n}|\mathbf{1})$ is the information (correlation) common to subsystems $X_1, ~X_2, \cdots$, and $X_n$. \\
4. It can assume negative values \cite{cerf1997, HOW2005, HOW2007, rio2011, gour2022}.

In the above, $I(X|\mathbf{1}) \equiv I(X)$, $I(X_1:X_2:X_3|\mathbf{1}) \equiv I(X_1:X_2:X_3)$, and so on. In multiparty systems, however, we prefer to keep the notation $I(X|\mathbf{1})$ rather than $I(X)$ to remind us the fact that either 
$I(X) \equiv I(X|\mathbf{1}) = -S(\mathbf{1}) + S(X)$ is analogous to conditional information or $I(X)$ is not a multiparty (total) correlation.  

\begin{center}
\begin{figure}[htb]
\subfloat[]{\includegraphics[width=1.5in, angle=0]{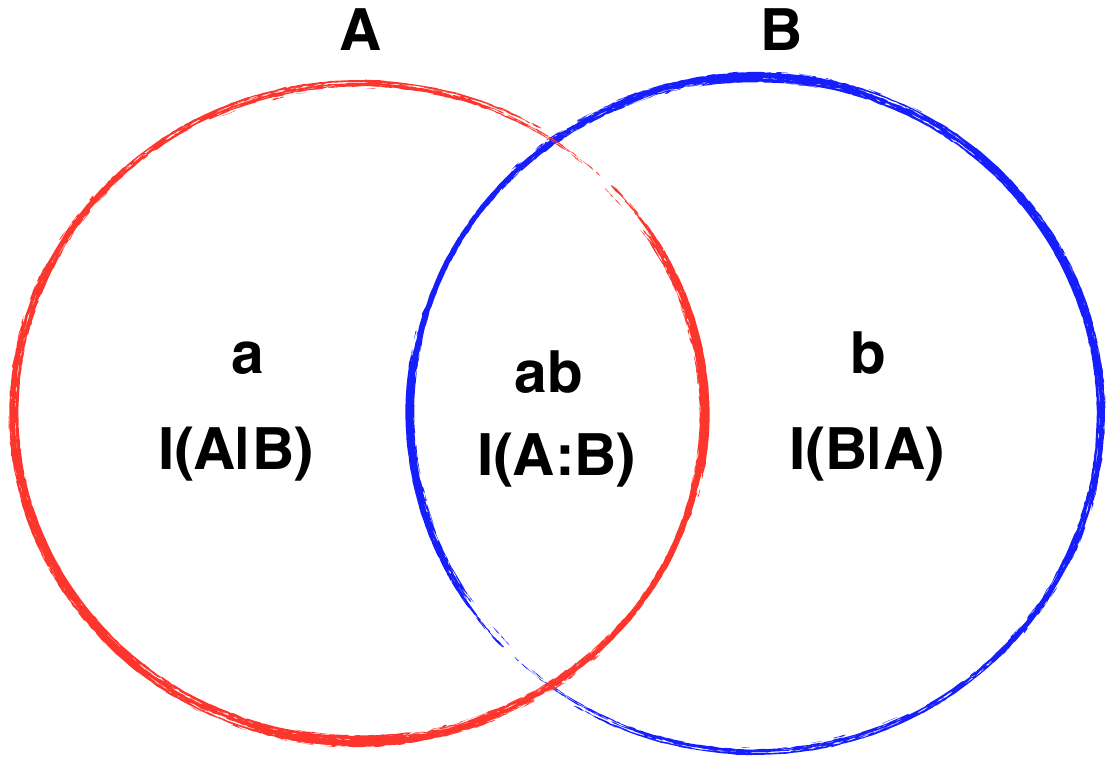}} \hspace{0.3cm}
\subfloat[]{\includegraphics[width=1.4in, angle=0]{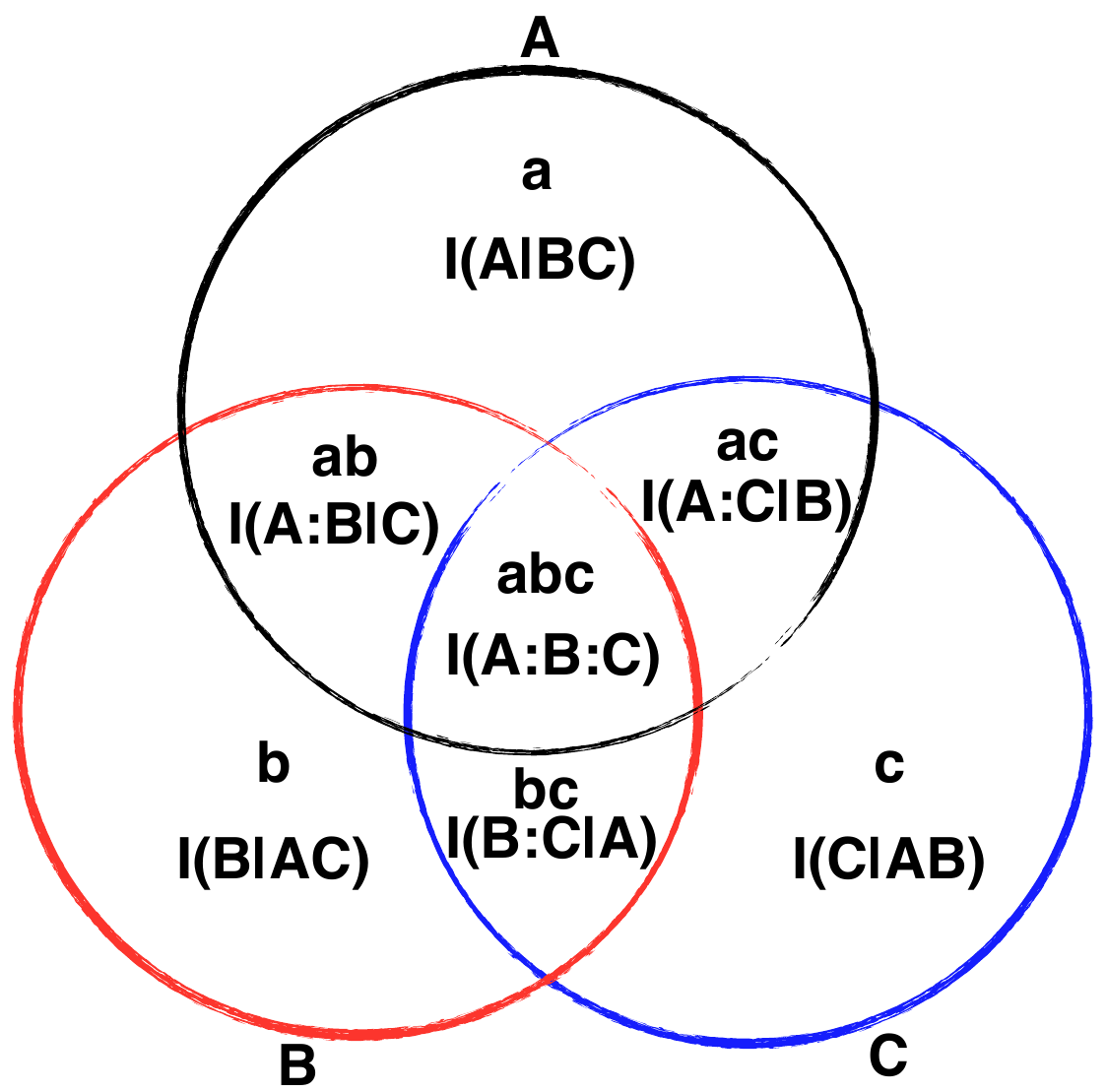}}
\caption{(a) Two-variable and (b) three-variable Venn diagrams with possible intersecting regions and generalized conditional mutual information. 
Here $S(X) \equiv I(X|\mathbf{1})$ is the information content of subsystem $X$, the information contained in subsystem $X$ but not in subsystem $Y$ is given by the conditional entropy $S(X|Y) = S(XY) - S(Y) = S(X) - I(X:Y) \equiv I(X|Y)$, and $I(X:Y)$ is the information between $X$ and $Y$. 
}
\label{fig1}
\end{figure}
\end{center}

The nontriviality in characterization of information and correlations begins to emerge with three-party system onwards. 
We illustrate the idea of GCMI using the tripartite system $ABC$ represented by density matrix $\rho_{ABC}$ [see Fig. \ref{fig1}(b)]. \\
1. Information in $A$ is $[a + ab + ac + abc] = S(\rho_A) \equiv I(A|\mathbf{1})$. \\
2. Information in $A$ but neither in $B$ nor in $C$ (i.e., information strictly contained in $A$) is
$[a] = - S(\rho_{BC}) + S(\rho_{ABC}) \equiv I(A|BC)$. \\
3. Information in $A$ (and possibly in $B$) but not in $C$ is $[a + ab]= - S(\rho_{C}) + S(\rho_{AC}) \equiv I(A|C)$. \\
4. Information common to $A$ and $B$ (and possibly in $C$) is $[ab + abc] = S(\rho_{A}) + S(\rho_{B}) - S(\rho_{AB}) \equiv I(A:B|\mathbf{1})$.  \\
5. Information common to $A$ and $B$ but not in $C$ is $[ab] = - S(\rho_{C}) + S(\rho_{AC}) + S(\rho_{BC}) - S(\rho_{ABC}) \equiv I(A:B|C)$. \\
6. Information common to $A$, $B$ and $C$ is $[abc] = S(\rho_{A}) + S(\rho_{B}) + S(\rho_{C}) - S(\rho_{AB}) - S(\rho_{AC}) - S(\rho_{BC}) + S(\rho_{ABC}) \equiv I(A:B:C|\mathbf{1})$. \\
7. Information in $ABC$ (as a single system) is $S(\rho_{ABC}) \equiv I(ABC|\mathbf{1})$. \\
Indeed, there are several possibilities. Different arrangements or configurations of subsystems yield, in general, different values of information or correlation. \\ 
8. Three-party total correlations amongst $A$, $B$ and $C$ are given by 
\begin{align}
& T_3(A:B:C) = S(\rho_{A}) + S(\rho_{B}) + S(\rho_{C}) - S(\rho_{ABC}) \nonumber \\
=& D(\rho_{ABC} \parallel \rho_A \otimes \rho_B \otimes \rho_C) \nonumber \\
=& I(A:BC) + I(B:C) \nonumber \\
=& I(A:B|C) + I(A:C|B) + I(B:C|A) + 2 I(A:B:C|\mathbf{1}), \nonumber
\end{align}
and
\begin{align}
& S_3(A:B:C) = S(\rho_{AB}) + S(\rho_{AC}) + S(\rho_{BC}) - 2S(\rho_{ABC}) \nonumber \\
=& I(A:BC) + I(B:C|A) \nonumber \\
=& I(A:B|C) + I(A:C|B) + I(B:C|A) + I(A:B:C|\mathbf{1}) \nonumber \\
=& S(\rho_{ABC}) - I(A|BC) - I(B|AC) - I(C|AB). \nonumber
\end{align}
%
%

\section{Family of MQMI}
\label{sec:mqmi}
The bipartite QMI $I(A:B) = S(A) + S(B) - S(AB)$ is the measure of total correlation (classical and quantum) between subsystems $A$ and $B$ \cite{groisman, luo2007}, satisfies the Araki-Lieb inequality $I(A:B) \le 2~ \text{min} \{S(\rho_A), S(\rho_B) \}$ \cite{wehrl1978}, is invariant under local unitary operations, and is nonincreasing under tracing out a subsystem.
Here we introduce a family of entropic functions constructed on $\rho_{X_1 \cdots X_n}$, and show below that they serve as $n$-party quantum mutual information. 

\subsection{Definition}
We define the $k^{th}~~(1 \le k \le n)$ multiparty quantum mutual information on $\rho_{X_1 \cdots X_n}$ as
\begin{align}
& M_k^{(n)}(X_1:X_2: \cdots :X_n) \nonumber \\
:=& \sum_{j_1 < \cdots < j_k \in [n]} S(X_{j_1} \cdots X_{j_k}) - \binom{n-1}{k-1} S(X_1 \cdots X_n). \label{qmi:k}
\end{align}
The superscript ``$n$'' in $M_k^{(n)}$ denotes the number of subsystems (single or composite) separated by colons, and the coefficients $\binom{n-1}{k-1}$ for given $n$ constitute the $n^{th}$-row of the Pascal's triangle.
We posit a few remarks here. First, the choice of definition in Eq.(\ref{qmi:k}) is motivated by the fact that the two well-established MQMI \cite{asu-qmi, cerf} in the literature are members of this family. Second, this multiparty quantum mutual information contains all two and more parties interactions, as described in \cite{asu-qmi}, and not just the $n$-party interaction $I(X_{1}: X_{2}: \cdots :X_{n}|\mathbf{1})$. Third, $M_1^{(2)}(X : Y) = I(X: Y) = I(X: Y | \mathbf{1})$.

For $p + q = n$, one can consider $p$-party versus $q$-party partitions of $\mathcal{X}$ 
such that
\begin{eqnarray}
M_p^{(n)} + M_q^{(n)} &=& \sum_{k \overline{k} \in \{ p|q ~partitions ~ of ~ \mathcal{X} \}} I \left(\rho_{k} : \rho_{\overline{k}} \right).
\label{qmi:pq}
\end{eqnarray}
See Appendix \ref{sec:relations} for more relations between $M_k^{(n)}$. 

The existence of several expressions for multiparty quantum mutual information can have profound consequences and diverse applications in quantum information theory and technology. 
As research in this area continues to evolve, the insights gained from multiple MQMI measures will be instrumental in unlocking the full potential of quantum information science. The following observations, for instance, merit attention.\\
1. $M_k^{(n)}$ are independent multiparty quantum mutual information. As these yield distinct values for an arbitrary state, the parties must make {\it a priori} choice about which multiparty correlation they want to consider. This {\it a priori} choice should be obvious (self-revealing) once their interpretations are known. \\
2. To maximize their correlation, parties should consider the quantity $M_k^{(n)}$ where $k=n/2$ (for even $n$) or $k=(n \pm 1)/2$ (for odd $n$). For minimal correlation, $M_1^{(n)}$ and $M_{n-1}^{(n)}$ are good candidates. \\
3. Corresponding to this family of multiparty quantum mutual information, one can introduce a family of multiparty quantum discord. Discord is the difference between {\em unmeasured} total correlation and {\em measured} total correlation present in a quantum state. \\
4. Information deviation due to a map or channel $\Phi$ (this may include noise) acting on a multiparty quantum state $\rho$ is $| {\cal Q}(\Phi(\rho)) - {\cal Q}(\rho) |$, where ${\cal Q}$ is either $M_k^{(n)}$, $M^{(n)}$, $C^{(n)}$, or generalized conditional mutual information. \\
5. Suppose the eavesdropper $E$ interacts with the system $\rho_{AB}$. Then, the desirable condition for the secure secret sharing of information between $A$ and $B$ is that the sum of correlations of $E$ with $A$ and $B$ should be minimal (refer to Fig. \ref{fig1}(b) with $E$ in place of $C$). That is,
\begin{align}
& I(A:E|B) + I(B:E|A) + I(A:B:E) \nonumber \\
=& M_2^{(3)} (A:B:E) - I(A:B|E) \nonumber \\
=& I(AB : E) \rightarrow 0. 
\end{align} 
A more stringent condition would be that each of $I(A:E|B)$, $I(B:E|A)$, and $I(A:B:E)$ is either zero or tends to zero. \\
6. $M_k^{(n)}$ together with $\sum_k c_k M_k^{(n)}$ can be used to distinguish quantum states.

The various equivalent expressions of two eminent MQMIs \cite{asu-qmi, cerf} are: 
\begin{align}
& T_n(X_1:X_2: \cdots :X_n) \equiv M_1^{(n)}(X_1:X_2: \cdots :X_n) \nonumber \\
=& \sum_{k=1}^n S(\rho_{X_k}) - S(\rho_{X_1 \cdots X_n}) \label{qmi:1ent} \\
=& D(\rho_{X_1 \cdots X_n} \parallel \rho_{X_1} \otimes \cdots \otimes \rho_{X_n})  
\label{qmi:1rel} \\
=& \sum_{k=1}^{n-1} I(X_k : X_{k+1} \cdots X_n) \label{qmi:1} \\
=& \sum_{j=2}^{n} (j-1) \sum_{k_1 < \cdots k_j \in [n]} I(X_{k_1} : \cdots :X_{k_j} | X_{k_{j+1}} \cdots X_{k_n}), \nonumber \\ \label{qmi:1regions}
\end{align}
and
\begin{align}
& S_n(X_1:X_2: \cdots :X_n) \equiv M_{n-1}^{(n)}(X_1:X_2: \cdots :X_n) \nonumber \\
=& \sum_{k=1}^n S(\rho_{X_1 \cdots X_{k-1}X_{k+1} \cdots X_n}) - (n-1)S(\rho_{X_1 \cdots X_n}) \label{qmi:n-1ent} \\
=& I(X_1 : X_{2} \cdots X_n) + \sum_{k=2}^{n-1}  I(X_k : X_{k+1} \cdots X_n | X_1 \cdots X_{k-1}) \label{qmi:n-1} \\
=& \sum_{j=2}^{n} ~ \sum_{k_1 < \cdots k_j \in [n]} I(X_{k_1} : \cdots :X_{k_j} | X_{k_{j+1}} \cdots X_{k_n}) \label{qmi:n-1regions} \\
=& S(\rho_{X_1 \cdots X_n}) - \sum_{k=1}^{n} I(X_k | \overline{X_k}). \label{qmi:n-1single}
\end{align}
Both $T_n$ and $S_n$ measure both the classical and the quantum correlations. 
$T_n$ and $S_n$ are referred to as ``total correlation'' \cite{groisman, watanabe} and ``dual total correlation'' \cite{han1975, han1978} respectively, quantum secrecy monotones \cite{cerf}, and multiparty quantum mutual information \cite{asu-qmi} from the information-theoretic point of view.

We further find that $M_k^{(n)}$, for fixed $k~~(1 \le k < n)$, is nondecreasing under discarding of any one party or grouping together any two parties (see Appendix \ref{sec:proof1}). That is,
\begin{align}
& M_k^{(n)}(X_1: \cdots :X_n) \ge  M_k^{(n-1)}(X_1: \cdots :X_{n-1}), \label{qmi:ineq1} \\
& M_k^{(n)}(X_1: \cdots :X_n) \ge  M_k^{(n-1)}(X_1: \cdots :X_{n-1}X_n). \label{qmi:ineq2}
\end{align}

\subsection{Interpretations}
{\it Interpretations of $T_n = M_1^{(n)}$.--} From Eq.(\ref{qmi:1rel}), $T_n$ is interpreted as the minimal relative entropy between $\rho_{X_1 \cdots X_n}$ and a product density matrix $\sigma_{X_1} \otimes \cdots \otimes \sigma_{X_n}$ (the minimum being attained when $\sigma_{X_k} = \rho_{X_k}$). From Eq.(\ref{qmi:1}), $T_n$ is the sum of decorrelation costs when the parties decorrelate themselves locally one by one from the rest using $I(X_k : X_{k+1} \cdots X_n)$ bits of randomness \cite{groisman}.  
From Eq.(\ref{qmi:1regions}), $T_n$ is the sum of $(j-1)$-times all $(j \ge 2)$-party interactions. 

{\it Interpretations of $S_n = M_{n-1}^{(n)}$.--} $S_n$ like $T_n$, from Eq.(\ref{qmi:n-1}), can be interpreted as the sum of decorrelation costs when the parties decorrelate themselves locally one by one from the rest using $I(X_k : X_{k+1} \cdots X_n | X_1 \cdots X_{k-1})$ bits of randomness. $S_n$, using Eq.(\ref{qmi:n-1regions}), is the sum of 
all interactions of two and more parties only once \cite{asu-qmi}. Equivalently, from Eq.(\ref{qmi:n-1single}), it is the entropy of the whole system less the sum of information in nonintersecting regions.

Nevertheless, while all $M_k^{(n)}$ can be viewed as the sum of two and more parties interactions, not all of them have straightforward physical or operational meaning like $M_1^{(n)}$ and $M_{n-1}^{(n)}$. Therefore, any operational interpretation of these mutual information would be appreciated.

\begin{table}[htb]
\begin{tabular}{cccccc}
\hline
\text{State} & $M_1^{(n)}$	& $M_2^{(n)}$		& $M_3^{(n)}$		& $M_4^{(n)}$		& $C^{(n)}$          \\ \hline 
$|gGHZ_2\rangle$                     & 2 $h(p)$  & 0  & $\times$  & $\times$  & 2 $h(p)$    \\ \hline 
$|gGHZ_3\rangle$                      & 3 $h(p)$  & 3 $h(p)$  & 0  & $\times$   & 0    \\ 
$|D_3^1\rangle$                      & 2.75489  & 2.75489  & 0  & $\times$  & 0 \\ 
$|\psi_{as}\rangle$                  & 4.75489   & 4.75489  & 0  & $\times$  & 0 \\ \hline 
$|gGHZ_4\rangle$                      & 4 $h(p)$   & 6 $h(p)$  & 4 $h(p)$  & 0 & 2 $h(p)$  \\ 
$|D_4^1\rangle$               & 3.24511 & 6  & 3.24511  & 0  & 0.490225 \\ 
$|D_4^2\rangle$               & 4  & 7.50978  & 4  & 0  & 0.490225 \\ 
$|C_4\rangle$                       & 4   & 10  & 4  & 0  & \bf{-2}   \\ 
$|HS_4\rangle$                       & 4   & 10.75489  & 4  & 0  & \bf{-2.75489}   \\ \hline 
$|gGHZ_5\rangle$                      & 5 $h(p)$   & 10 $h(p)$  & 10 $h(p)$  & 5 $h(p)$ & 0  \\ 
$|D_5^1\rangle$               & 3.60964 & 9.70951  & 9.70951  & 3.60964  & 0 \\ 
$|D_5^2\rangle$               & 4.85475 & 12.95462  & 12.95462  & 4.85475  & 0 \\ \hline
\end{tabular}
\caption{Values of multiparty quantum mutual information $M_k^{(n)}$ and common information $C^{(n)}$ of generalized Greenberger-Horne-Zeilinger states $|gGHZ_n\rangle=\sqrt{p}~|0\rangle^{\otimes n} + e^{i\phi} \sqrt{1-p}~ |1\rangle^{\otimes n}$, 
Dicke states $|D_n^r\rangle=\frac{1}{\sqrt{\binom{n}{r}}}\sum_{{\cal P}}{\cal P}[|0\rangle^{\otimes n-r}|1\rangle^{\otimes r}]$, 
three-qutrit totally antisymmetric state $|\psi_{as}\rangle=
\frac{1}{\sqrt{6}}(|123\rangle-|132\rangle+|231\rangle-|213\rangle+|312\rangle-|321\rangle)$, 
four-qubit cluster state $|C_4\rangle=\frac{1}{2}(|0000\rangle+|0011\rangle+|1100\rangle-|1111\rangle)$, 
and $|HS_4\rangle = \frac{1}{\sqrt{6}} (|0011\rangle + |1100\rangle + \omega (|1010\rangle + |0101\rangle) + \omega^2 (|1001\rangle + |0110\rangle))$. Here $h(p) := - p \log_2 p -  (1-p) \log_2 (1-p)$ is the binary entropy, $\omega = e^{\frac{2\pi i}{3}}$, and ``$\times$'' stands for {\it not applicable}. Note that  $C^{(n)}$ can be negative and $C^{(n=odd)}$ vanishes for pure states.}
\label{table-qmi}
\end{table}

\subsection{Properties}
Here we discuss a number of useful properties that $M_k^{(n)}$ satisfy. \\
1. $M_k^{(n)}$ is invariant under local unitary operations because von Neumann entropy is invariant under unitary operations. \\ 
2. $M_1^{(n)} = T_n$ and $M_{n-1}^{(n)} = S_n$ satisfy the following properties \cite{cerf, asu-qmi}: (P1--P9) above, (P10) $S_n = T_n$ for pure states, and (P11) $S_n(\rho_{X_1X_2 \cdots X_n}) \le T_n(\rho_{X_1X_2 \cdots X_n}) + 2S(\rho_{X_1X_2 \cdots X_n})$. \\
3. $M_n^{(n)} = 0$ (as expected) because $(X_1X_2 \cdots X_n)$ as a single system has no mutual information. \\
4. Suppose $1 \le p, q \le n-1$ such that $p \neq q$ and $p + q = n$. 
Then $M_p^{(n)} = M_q^{(n)}$ for pure states and hence can be called ``dual'' to each other. 
We envisage from Table \ref{table-qmi} that for pure states the profile of $M_k^{(n)}$ versus $k$ is analogous to a truncated Gaussian (a point or a straight line being the particular case). \\
5. $M_k^{(n)}$ 
satisfy (P1) symmetry, (P3) vanishing on product states, (P6) additivity and (P7) continuity. These properties are also satisfied by any quantity which is a linear combination of $M_k^{(n)}$. In particular,
\begin{align}
& M^{(n)}(X_1:X_2: \cdots :X_n) \nonumber \\
=& \sum_{k=1}^{n} \lambda_k  M_k^{(n)}(X_1:X_2: \cdots :X_n), 
\label{qmi:comb}
\end{align}   
where $\lambda_k \ge 0$ and $\sum_k \lambda_k = 1$. We conjecture below that $M_k^{(n)}$ and 
$M^{(n)}$ are plausible candiadtes for secrecy monotones.
Another important linear combination of $M_k^{(n)}$ is 
\begin{align}
& C^{(n)}(X_1:X_2: \cdots :X_n) \nonumber \\
=& \sum_{k=1}^{n} (-1)^{k+1} M_k^{(n)}(X_1:X_2: \cdots :X_n), \label{qmi:common1} \\
=& I(X_1:X_2: \cdots :X_n|\mathbf{1}). \nonumber
\end{align}   
$C^{(n)}$ is the information (correlation) common to all $X_k$s. It can, however, be negative (see Table \ref{table-qmi}). 
Moreover, it vanishes identically for pure {\it odd}-party quantum states: 
\begin{eqnarray}
C^{(n = \text{odd})} = \sum_{k = \text{odd}} M_k^{(n)} - \sum_{k = \text{even}} M_k^{(n)} = 0.
\label{qmi:common2}
\end{eqnarray}
6. For pure states, $C^{(n)} \le M_k^{(n)} = M_{n-k}^{(n)}$. For mixed states, we expect them to obey the inequality: 
\begin{align}
& C^{(n)} \le \left( M_1^{(n)} \approx M_{n-1}^{(n)} \right) \nonumber \\
&< \left( M_2^{(n)} \approx M_{n-2}^{(n)} \right) < \cdots < M_{n/2}^{(n)}.
\end{align}
We also surmise that for $k_1 < k_2 \le n/2$ and $c \ge c_{k_2,k_1} = \frac{k_2 \binom{n}{k_2}}{k_1 \binom{n}{k_1}}$,
\begin{align}
c M_{k_1}^{(n)} \ge M_{k_2}^{(n)}.
\end{align}
7. $M_k^{(n)}(X_1:X_2: \cdots :X_n)$ in Eq.(\ref{qmi:k}) is semipositive. \\
{\it Proof.}
It is obvious for pure states because $S(X_1X_2 \cdots X_n)$ vanishes identically. 
For mixed states, the semipositivity of $M_1^{(n)}$, $M_{n-1}^{(n)}$, and $M_{k=n/2}^{(n)}$ (for even $n$) follows from Eqs.(\ref{qmi:1rel}, \ref{qmi:1}), Eq.(\ref{qmi:n-1}), and Eq.(\ref{qmi:pq}) respectively. In general, $M_k^{(n)}$ is semipositive because it is nondecreasing under discarding a subsystem: $M_k^{(n)} \ge M_k^{(n-1)} \ge \cdots \ge M_k^{(k)} =0$ [see Eq.(\ref{qmi:ineq1})]. 
$M_k^{(n)}$ can also be shown nonnegative using the subadditivity and the strong subadditivity of von Neumann entropy (see Appendix \ref{sec:proof2}). 
The idea is to eliminate $S(X_1X_2 \cdots X_n)$ terms. 
This can be achieved by grouping together (repeatedly) two appropriate entropy terms of small number of parties to obtain an entropy term having greater number of parties. One should, however, take care that while grouping no entropy term, except $S(X_1X_2 \cdots X_n)$, appears twice or more. 
\hfill $\blacksquare$

Thus, $M_k^{(n)}$ in Eq.(\ref{qmi:k}) and $M^{(n)}$ in Eq.(\ref{qmi:comb}) satisfy a number of useful properties: symmetry, semipositivity, vanishing on product states, additivity, and continuity. Hence, they constitute a family of multiparty quantum mutual information.

\subsection{Conjecture and Remark}
Secrecy monotones quantify the amount of {\em secret correlation} shared by the parties of a multipartite system. They are useful in the study of quantum (as well as classical) cryptography.
$M_1^{(n)} = T_n$ and $M_{n-1}^{(n)} = S_n$ satisfy properties (P4, P5) and (P8, P9), and qualify for secrecy monotones \cite{cerf}.
We speculate (see Appendix \ref{sec:proof3} for argument) that $M_k^{(n)}~~(k = 2, 3, \cdots ,n-2)$ in Eq.(\ref{qmi:k}) and $M^{(n)}$ in Eq.(\ref{qmi:comb}) also meet the criteria (P4, P5) and (P8, P9) for being considered as measures of secrecy monotone. 
%

In addition to these $n$-party symmetric monotones, we also have other monotones $M_k^{(m)}~~(1 \le k \le m < n)$ on $n$-party quantum states $\rho_{X_1 \cdots X_n}$ which can be obtained by grouping together any two or more of the $n$-parties. All these monotones, however, are not all linearly independent.

\section{Conclusions}
\label{sec:conclusion}
In this study, we have conducted a comprehensive analysis of entropy-based information within multiparty systems. Firstly, we introduced the concept of generalized conditional mutual information. Next, we presented a family of multiparty quantum mutual information, which is anticipated to significantly contribute to fundamental research in quantum information theory. This advancement is expected to enhance our comprehension of classical, quantum, and total correlations, and consequences thereof. Notably, this framework includes the two well-established multiparty quantum mutual information measures.

While various interpretations exist for \(M_1^{(n)}\) and \(M_{n-1}^{(n)}\), the interpretation of other measures \(M_k^{(n)}\) remains unclear. Therefore, some operational interpretations of these mutual information measures would be beneficial.
Additionally, we conjecture that the remaining measures \(M_k^{(n)}\) and \(M^{(n)} = \sum_k \lambda_k M_k^{(n)}\), where \(\lambda_k \ge 0 ~\text{and}~ \sum_k \lambda_k =1\), will exhibit monotonicity under local operations and classical communication, thereby qualifying as secrecy monotones. We also posit that our formalism will be instrumental in characterizing measures of multiparty nonclassical correlations.

\begin{widetext}
\appendix 
\section{Relations between $M_k^{(n)}$}
\label{sec:relations}
\begin{enumerate}
\item The recurrence relation(s) for $T_n$ and $S_n$ are as follows:
\begin{eqnarray}
T_n(X_1:X_2: \cdots :X_n) 
&=& T_{n-1}(X_{k_1}:X_{k_2}: \cdots :X_{k_{n-1}}) + I (X_{k_1} \cdots X_{k_{n-1}} : X_{k_n}),
\label{qmi:1rec1} \\ 
T_n(X_1:X_2: \cdots :X_n)  &=& T_{n-1}(X_{k_1}X_{k_2}: X_{k_3}: \cdots :X_{k_n}) + I (X_{k_1}:X_{k_2}),
\label{qmi:1rec2}
\end{eqnarray}
and 
\begin{eqnarray}
S_n(X_1:X_2: \cdots :X_n) 
&=& S_{n-1}(X_{k_1}X_{k_2}: X_{k_3}: \cdots :X_{k_n}) + I (X_{k_1}:X_{k_2} | X_{k_3} \cdots X_{k_n}).
\label{qmi:n-1rec}
\end{eqnarray}
where $X_{k_j}$ belongs to and exhaust the set $\mathcal{X}$. It is evident that the recurrence relation is not unique. For the choice of $\{X_{k_{n-j}} = X_{j+1} \}_{j=0}^{n-1}$, Eq.(\ref{qmi:1rec1}) and Eq.(\ref{qmi:1rec2}) separately yields Eq.(\ref{qmi:1}), and Eq.(\ref{qmi:n-1rec}) yields Eq.(\ref{qmi:n-1}). 


\item Relations between $M_{k}^{(n_1)}$ and $M_{k+1}^{(n_2)}$.
\begin{align}
& M_{2}^{(n)} + \sum_{j < k \in [n]}  I(X_j : X_k) = (n-1) M_{1}^{(n)}, \\
& (k +1) M_{k+1}^{(n)} + \sum_{j_1 < \cdots < j_k \in [n]} ~ \sum_{i (\neq j_1 \neq \cdots \neq j_k)} I \left( X_{j_1} \cdots X_{j_k} : X_i \right) 
= (n-k) M_k^{(n)} + \left(1 - \frac{k}{n} \right) \binom{n}{k} M_1^{(n)}, \label{qmi:k2k+1} \\
& M_{n-1}^{(n)}(X_1: \cdots : X_n) = M_{n-2}^{(n-1)}(X_1: \cdots : X_{n-1}) + \sum^{n-1}_{k=1} \left( S_{\overline{X_k}} + S_{\overline{X_n}} - S_{X_1 \cdots X_n} - S_{\overline{X_k X_n}} \right).
\end{align}
%
\end{enumerate}

\section{Proof of Eqs.(\ref{qmi:ineq1}, \ref{qmi:ineq2})}
\label{sec:proof1}
Here we show that $M_k^{(n)}$, for fixed $k~~(1 \le k < n)$, is nondecreasing under discarding of any one party or grouping together any two parties. That is,
\begin{eqnarray}
M_k^{(n)}(X_1:X_2: \cdots :X_n) &\ge & M_k^{(n-1)}(X_1:X_2: \cdots :X_{n-1}), 
\\
M_k^{(n)}(X_1:X_2: \cdots :X_n) &\ge & M_k^{(n-1)}(X_1:X_2: \cdots :X_{n-1}X_n). 
\end{eqnarray}

Above inequalities follow from the definition of $M_k^{(n)}$ and the entropic inequalities 
$S(X) + S(Y) \ge S(XY)$ (subadditivity) and $S(XY) + S(YZ) \ge S(Y) + S(XYZ)$ (strong subadditivity), as shown below.
First, we show that $M_k^{(n)}$ is nondecreasing under discarding a subsystem.
\begin{eqnarray}
M_1^{(n)} (X_1:X_2: \cdots :X_n) &=& \sum_{k=1}^n S_k - S_{12 \cdots n} \nonumber \\
&=& \sum_{k=1}^{n-1} S_k - S_{12 \cdots (n-1)} + \left( S_{12 \cdots (n-1)} + S_n - S_{12 \cdots n} \right) \nonumber \\
&=& M_1^{(n-1)} (X_1:X_2: \cdots :X_{n-1}) + I(X_1X_2 \cdots X_{n-1} : X_n | \mathbf{1}) \nonumber \\
&=& M_1^{(n-1)} (X_1:X_2: \cdots :X_{n-1}) + M_1^{(2)}(X_1X_2 \cdots X_{n-1} : X_n), \nonumber \\
& \ge & M_1^{(n-1)} (X_1:X_2: \cdots :X_{n-1}), \nonumber 
\end{eqnarray}
\begin{eqnarray}
M_2^{(n)} (X_1:X_2: \cdots :X_n) &=& \sum_{j<k \in [n]} S_{jk} - (n-1)S_{12 \cdots n} \nonumber \\
&=& \sum_{j<k \in [n-1]} S_{jk} - (n-2)S_{12 \cdots (n-1)} \nonumber \\
&&+ \left( \sum_{j = 1}^{n-1} S_{jn} + (n-2)S_{12 \cdots (n-1)} - (n-1)S_{12 \cdots n} \right) \nonumber \\
&\ge & M_2^{(n-1)} (X_1:X_2: \cdots :X_{n-1}) +  \left( \sum_{j = 1}^{n-2} S_{j} + S_{(n-1)n} - S_{12 \cdots n} \right) \nonumber \\
&\ge & M_2^{(n-1)} (X_1:X_2: \cdots :X_{n-1}) + M_1^{(n-1)}(X_1:X_2: \cdots X_{n-2}: X_{n-1} X_n), \nonumber \\ 
&\ge & M_2^{(n-1)} (X_1:X_2: \cdots :X_{n-1}), \nonumber 
\end{eqnarray}
\begin{eqnarray}
M_3^{(n)} (X_1:X_2: \cdots :X_n) &=& \sum_{i<j<k \in [n]} S_{ijk} - \binom{n-1}{2}S_{12 \cdots n} \nonumber \\
&=& \sum_{i<j<k \in [n-1]} S_{ijk} - \binom{n-2}{2} S_{12 \cdots (n-1)} \nonumber \\
&&+ \left( \sum_{i<j \in [n-1]} S_{ijn} + \binom{n-2}{2} S_{12 \cdots (n-1)} - \binom{n-1}{2} S_{12 \cdots n} \right) \nonumber \\
&\ge & M_3^{(n-1)} (X_1:X_2: \cdots :X_{n-1}) +  \left( \sum_{j = 2}^{n-1} S_{1jn} + \sum_{i = 2}^{n-2} \sum_{j = i+1}^{n-1} S_{ij} - (n-2) S_{12 \cdots n} \right) \nonumber \\
&\ge & M_3^{(n-1)} (X_1:X_2: \cdots :X_{n-1}). \nonumber
\end{eqnarray}
Similarly, one can show that $M_{k>3}^{(n)}(X_1:X_2: \cdots :X_n) \ge M_{k>3}^{(n-1)}(X_1:X_2: \cdots :X_{n-1})$.
Next, we show that $M_k^{(n)}$ is nondecreasing under grouping together two parties, specifically for $n=4$. 
\begin{eqnarray}
M_1^{(4)} (X_1:X_2: X_3 :X_4) &=& (S_1 + S_2 + S_3 + S_4) - S_{1234} \nonumber \\
&=& (S_1 + S_2 + S_{34} - S_{1234}) + (S_3 + S_4 - S_{34}) \nonumber \\
&=& M_1^{(3)} (X_1:X_2: X_3 X_4) + M_1^{(2)} (X_3: X_4), \nonumber \\ 
&\ge & M_1^{(3)} (X_1:X_2: X_3 X_4), \nonumber 
\end{eqnarray}
\begin{eqnarray}
M_2^{(4)} (X_1:X_2: X_3 :X_4) &=& S_{12} + S_{13} + S_{14} + S_{23} + S_{24} + S_{34} - 3S_{1234} \nonumber \\
&=& (S_{12} + S_{134} + S_{234} - 2S_{1234}) \nonumber \\
&&+ (S_{13} + S_{14} + S_{23} + S_{24} + S_{34} - S_{134} - S_{234} - S_{1234}) \nonumber \\
&\ge & M_2^{(3)} (X_1:X_2: X_3 X_4) + (S_1 + S_2 + S_{34} - S_{1234}), \nonumber \\
&= & M_2^{(3)} (X_1:X_2: X_3 X_4) + M_1^{(3)} (X_1 : X_2 : X_3 X_4), \nonumber \\
&\ge & M_2^{(3)} (X_1:X_2: X_3 X_4), \nonumber 
\end{eqnarray}
\begin{eqnarray}
M_3^{(4)} (X_1:X_2: X_3 :X_4) &=& S_{123} + S_{124} + S_{134} + S_{234} - 3S_{1234} \nonumber \\
& \ge & 0 = M_3^{(3)} (X_1:X_2: X_3 X_4). \nonumber
\end{eqnarray}
\hfill $\blacksquare$

\section{Semipositivity of $M_k^{(n)}$}
\label{sec:proof2}
Here we illustrate the nonnegativity of $M_k^{(n)}$ for $n=5$.
\begin{eqnarray}
M_1^{(5)} &=& (S_1 + S_2 + S_3 + S_4 + S_5) - S_{12345} \ge S_{12345} - S_{12345} = 0. \nonumber \\ \nonumber \\
M_2^{(5)} &=& S_{12} + S_{13} + S_{14} + S_{15} + S_{23} + S_{24} + S_{25} + S_{34} + S_{35} + S_{45} - 4S_{12345} \nonumber \\
&=& (S_{12} + S_{34}) + S_{15} + (S_{13} + S_{45}) + S_{24} + (S_{14} + S_{23}) + S_{25} + S_{35} - 4S_{12345} \nonumber \\  
&\ge & (S_{1234} + S_{15}) + (S_{1345} + S_{24}) + (S_{1234} + S_{25}) + S_{35} - 4S_{12345} \nonumber \\  
&\ge & (S_{12345} + S_{1}) + (S_{12345} + S_{4}) + (S_{12345} + S_{2}) + S_{35} - 4S_{12345} \nonumber \\  
&= & (S_{1} + S_{2} + S_{4}) + S_{35} - S_{12345} \nonumber \\  
&\ge & (S_{124} + S_{35}) - S_{12345} \nonumber \\  
&\ge & S_{12345} - S_{12345} = 0. \nonumber \\ \nonumber \\
M_3^{(5)} &=& S_{123} + S_{124} + S_{125} + S_{134} + S_{135} + S_{145} + S_{234} + S_{235} + S_{245} + S_{345} - 6S_{12345} \nonumber \\
&=& (S_{123} + S_{145}) + (S_{124} + S_{235}) + (S_{135} + S_{234}) + (S_{134} + S_{245}) + (S_{125} + S_{345}) - 6S_{12345} \nonumber \\
&\ge & (S_{1} + S_{12345}) + (S_{2} + S_{12345}) + (S_{3} + S_{12345}) + (S_{4} + S_{12345}) + (S_{5} + S_{12345}) - 6S_{12345} \nonumber \\
&=& (S_1 + S_2 + S_3 + S_4 + S_5) - S_{12345} \ge 0. \nonumber \\ \nonumber \\
M_4^{(5)} &=& (S_{1234} + S_{1235}) + (S_{1245} + S_{1345}) + S_{2345} - 4S_{12345} \nonumber \\
&\ge & (S_{123} + S_{12345}) + (S_{145} + S_{12345}) + S_{2345} - 4S_{12345} \nonumber \\
&\ge & (S_{123} + S_{145}) + S_{2345} - 2S_{12345} \nonumber \\
&\ge & (S_{1} + S_{12345}) + S_{2345} - 2S_{12345} \nonumber \\
&= & (S_{1} + S_{2345}) - S_{12345} \nonumber \\
&\ge & S_{12345} - S_{12345} = 0. \nonumber
\end{eqnarray}

Alternatively, one can also endeavor to obtain a recurrence relation for $M_k^{(n)}$ which expresses it as a positive sum of bipartite mutual information $I(A:B)$ and conditional mutual information $I(A:B|C)$. Then, the semipositivity of $M_k^{(n)}$ is trivial. We, however, note that obtaining the recurrence relation is neither unique [see Eqs.(\ref{qmi:1rec1}, \ref{qmi:1rec2}, \ref{qmi:n-1rec})] nor easy. For example,
\begin{eqnarray}
& M_2^{(n)}(X_1:X_2: \cdots :X_n) = M_2^{(n-1)}(X_1:X_2: \cdots :X_{n-1}) + M_1^{(n-1)}(X_1:X_2: \cdots :X_{n-1}) \nonumber \\
& +  I(X_n: X_{2} \cdots X_{n-1} | X_1) + \sum_{j=2}^{n-1} I(X_n: X_1 \cdots X_{j-1} X_{j+1} \cdots X_{n-1} | X_j).
\label{qmi:2rec}
\end{eqnarray}
\hfill $\blacksquare$

\section{Argument for Secrecy Monotones of $M_k^{(n)}$}
\label{sec:proof3}
Our argument for the speculation that $M_k^{(n)}$ meet the criteria of secrecy monotones is as follows. Consider a five-party quantum system $\mathcal{X} = \{X_1, \cdots, X_5 \}$, for example. We know that the von Neumann entropy is invariant under unitary transformations including the permutation or particle exchange operator. That is, $S_{12345} = S_{13245} = S_{23145}$, etc. Then
\begin{align}
& M_2^{(5)} (X_1 : X_2: \cdots : X_5) = S_{12} + S_{13} + S_{14} + S_{15} + S_{23} + S_{24} + S_{25} + S_{34} + S_{35} + S_{45} - 4S_{12345} \nonumber \\
=& \left(S_{12} + \log_2(d_3d_4d_5) - S_{12345} \right) + \left(S_{13} + \log_2(d_2d_4d_5) - S_{13245} \right) + \cdots \nonumber \\
+& \left(S_{45} + \log_2(d_1d_2d_3) - S_{12345} \right) 
+ 6S_{12345} - \log_2(d_1d_2d_3d_4d_5)^6 \nonumber \\
=& D(\rho_{12345} || \rho_{12} \otimes I_{345}) + D(\rho_{13245} || \rho_{13} \otimes I_{245}) + \cdots + D(\rho_{12345} || I_{123} \otimes \rho_{45}) + 6S_{12345} - \log_2(d_1d_2d_3d_4d_5)^6, \nonumber 
\end{align}
and
\begin{align}
& M_3^{(5)} (X_1 : X_2: \cdots : X_5) = S_{123} + S_{124} + S_{125} + S_{134} + S_{135} + S_{145} + S_{234} + S_{235} + S_{245} + S_{345} - 6S_{12345} \nonumber \\
=& \left(S_{123} + \log_2(d_4d_5) - S_{12345} \right) + \left(S_{124} + \log_2(d_3d_5) - S_{12435} \right) + \cdots \nonumber \\
+& \left(S_{345} + \log_2(d_1d_2) - S_{12345} \right) 
+ 4S_{12345} - \log_2(d_1d_2d_3d_4d_5)^4 \nonumber \\
=& D(\rho_{12345} || \rho_{123} \otimes I_{45}) + D(\rho_{12435} || \rho_{124} \otimes I_{35}) + \cdots + D(\rho_{12345} || I_{12} \otimes \rho_{345}) + 4S_{12345} - \log_2(d_1d_2d_3d_4d_5)^4. \nonumber 
\end{align}

While $D(\rho_{AB} || \rho_{A} \otimes \rho_{B})$ is always well-behaved, $D(\rho_{AB} || \rho_{B} \otimes \rho_{A})$ is not in general due to the support condition. Therefore, in the second steps above, the parties (subsystems) have been rearranged beforehand to resolve the support condition of QRE.
Note that the operations such as local quantum operations and local measurements and public classical communication can be regarded as local positive maps \cite{cerf}.
Because $D(\rho || \sigma) \ge D(\Phi(\rho) || \Phi(\sigma))$ for any local positive map $\Phi$, it implies that $M_k^{(n)}$ are monotonic under local quantum operations and classical communications.  
\hfill $\blacksquare$

\end{widetext}

\begin{thebibliography}{99}%
\bibitem{thomas} T. Cover and J. Thomas, {\em Elements of Information Theory} (John Wiley \& Sons, 1991).

\bibitem{nielsen} M. A. Nielsen and I. L. Chuang, {\it Quantum Computation and Quantum Information} (Cambridge University Press, 2000).

\bibitem{wilde-qit} M. M. Wilde, {\it Quantum Information Theory} (Cambridge University Press, 2013).

\bibitem{preskill} J. Preskill, {\it Lecture Notes for Physics 229: Quantum Information and Computation} (CreateSpace Independent Publishing Platform, 2015). 

\bibitem{groisman} B. Groisman, S. Popescu, and A. Winter, {\it Quantum, classical, and total amount of correlations in a quantum state}, \href{https://doi.org/10.1103/PhysRevA.72.032317} {Phys. Rev. A {\bf 72}, 032317 (2005)}.

\bibitem{luo2007} N. Li and S. Luo, {\it Total versus quantum corrrelations in quantum states}, \href{https://doi.org/10.1103/PhysRevA.76.032327} {Phys. Rev. A {\bf 76}, 032327 (2007)}.

\bibitem{bennett} C. H. Bennett and P. W. Shor, {\it Quantum Channel Capacities}, \href{https://doi.org/10.1126/science.1092381} {Science {\bf 303}, 1784-1787 (2004)}.
 
\bibitem{holevo} A. S. Holevo, {\it Quantum channel capacities}, \href{https://doi.org/10.1070/QEL17285} {Quantum Electron. {\bf 50}, 440 (2020)}.

\bibitem{biamonte} J. Biamonte, P. Wittek, N. Pancotti, P. Rebentrost, N. Wiebe, and S. Lloyd, {\it Quantum machine learning}, \href{https://doi.org/10.1038/nature23474} {Nature {\bf 549}, 195-202 (2017)}.
 
\bibitem{carleo} G. Carleo, I. Cirac, K. Cranmer, L. Daudet, M. Schuld, N. Tishby, L. Vogt-Maranto, and L. Zdeborov\'a, {\it Machine learning and the physical sciences}, \href{https://doi.org/10.1103/RevModPhys.91.045002} {Rev. Mod. Phys. {\bf 91}, 045002 (2019)}.

\bibitem{tomasi} G. D. Tomasi, S. Bera, J. H. Bardarson, and F. Pollmann, {\it Quantum Mutual Information as a Probe for Many-Body Localization}, \href{https://doi.org/10.1103/PhysRevLett.118.016804} {Phys. Rev. Lett. {\bf 118}, 016804 (2017)}.

\bibitem{chisholm} D. A. Chisholm, L. Innocenti, and G. M. Palma, {\it Importance of using the averaged mutual information when quantifying quantum objectivity}, \href{https://doi.org/10.1103/PhysRevA.110.012218} {Phys. Rev. A {\bf 110}, 012218 (2024)}.

\bibitem{bengtsson} I. Bengtsson and K. \.{Z}yczkowski, {\it Geometry of Quantum States: An Introduction to Quantum Entanglement} (Cambridge University Press, 2006).

\bibitem{HHHH2009} R. Horodecki, P. Horodecki, M. Horodecki, and K. Horodecki, {\it Quantum entanglement}, \href{https://doi.org/10.1103/RevModPhys.81.865} {Rev. Mod. Phys. {\bf 81}, 865 (2009)}.

\bibitem{modi2012} K. Modi, A. Brodutch, H. Cable, T. Paterek, and V. Vedral, {\it The classical-quantum boundary for correlations: Discord and related measures}, \href{https://doi.org/10.1103/RevModPhys.84.1655} {Rev. Mod. Phys. {\bf 84}, 1655 (2012)}.

\bibitem{bera2018} A. Bera, T. Das, D. Sadhukhan, S. S. Roy, A. Sen(De), and U. Sen, {\it Quantum discord and its allies: a review of recent progress}, \href{https://doi.org/10.1088/1361-6633/aa872f} {Rep. Prog. Phys. {\bf 81}, 024001 (2018)}.

\bibitem{watanabe} S. Watanabe, {\it Information theoretical analysis of multivariate correlation}, \href{https://doi.org/10.1147/rd.41.0066} {IBM Journal of Research and Development {\bf 4(1)}, 66-81 (1960)}. 

\bibitem{han1975} T. S. Han, {\it Linear dependence structure of the entropy space}, \href{https://doi.org/10.1016/S0019-9958(75)80004-0} {Inform. Contr. {\bf 29}, 337-368 (1975)}. 

\bibitem{han1978} T. S. Han, {\it Nonnegative Entropy Measures of Multivariate
Symmetric Correlations}, \href{https://doi.org/10.1016/S0019-9958(78)90275-9} {Inform. Contr. {\bf 36}, 133-156 (1978)}.   

\bibitem{walczak} Z. Walczak, {\it Total correlations and mutual information}, \href{https://doi.org/10.1016/j.physleta.2009.03.047} {Phys. Lett. A {\bf 373}, 1818-1822 (2009)}.

\bibitem{giorgi2011} G. L. Giorgi, B. Bellomo, F. Galve, and R. Zambrini, {\it Genuine quantum and classical correlations in multipartite systems}, \href{https://doi.org/10.1103/PhysRevLett.107.190501} {Phys. Rev. Lett. {\bf 107}, 190501 (2011)}.

\bibitem{asu-qmi} A. Kumar, {\it Multiparty quantum mutual information: An alternative definition}, \href{https://doi.org/10.1103/PhysRevA.96.012332}{Phys. Rev. A {\bf 96}, 012332 (2017)}.

\bibitem{sazim} Sk Sazim and P. Agrawal, {\it Quantum mutual information and quantumness vectors for multiqubit systems}, \href{https://doi.org/10.1007/s11128-020-02718-1} {Quantum Inf. Process. {\bf 19}, 216 (2020)}.

\bibitem{wehrl1978} A. Wehrl, {\it General properties of entropy}, \href{https://doi.org/10.1103/RevModPhys.50.221} {Rev. Mod. Phys. {\bf 50}, 221 (1978)}.

\bibitem{ruskai} M. B. Ruskai, {\it Inequalities for quantum entropy: A review with conditions for equality}, \href{https://doi.org/10.1063/1.1497701} {J. Math. Phys. {\bf 43}, 4358-4375 (2002)}.

\bibitem{hermes} A. \"{M}uller-Hermes and D. Reeb, {\it Monotonicity of the Quantum Relative Entropy Under Positive Maps}, \href{https://doi.org/10.1007/s00023-017-0550-9}{Ann. Henri Poincar\'{e} {\bf 18}, 1777-1788 (2017)}.

\bibitem{cerf} N. J. Cerf, S. Massar, and S. Schneider, {\it Multipartite classical and quantum secrecy monotones}, \href{https://doi.org/10.1103/PhysRevA.66.042309}{Phys. Rev. A {\bf 66}, 042309 (2002)}.

\bibitem{HHH2000} M. Horodecki, P. Horodecki, and R. Horodecki, {\it Limits for Entanglement Measures}, \href{https://doi.org/10.1103/PhysRevLett.84.2014} {Phys. Rev. Lett. {\bf 84}, 2014 (2000)}.

\bibitem{cerf1997} N. J. Cerf and C. Adami, {\it Negative entropy and information in quantum mechanics}, \href{https://doi.org/10.1103/PhysRevLett.79.5194} {Phys. Rev. Lett. {\bf 79}, 5194-5197 (1997)}.

\bibitem{HOW2005} M. Horodecki, J. Oppenheim, and A. Winter, {\it Partial quantum information}, \href{https://doi.org/10.1038/nature03909} {Nature {\bf 436}, 673-676 (2005)}.

\bibitem{HOW2007} M. Horodecki, J. Oppenheim, and A. Winter, {\it Quantum State Merging and Negative Information}, \href{https://doi.org/10.1007/s00220-006-0118-x} {Commun. Math. Phys. {\bf 269}, 107-136 (2007)}. 

\bibitem{rio2011} L. del Rio, J. Aberg, R. Renner, O. Dahlsten, and V. Vedral,  {\it The thermodynamic meaning of negative entropy}, \href{https://doi.org/10.1038/nature10395} {Nature {\bf 474}, 61-63 (2011)}.

\bibitem{gour2022} G. Gour, M. M. Wilde, S. Brandsen, and I. J. Geng, {\it Inevitability of knowing less than nothing}, \href{https://doi.org/10.48550/arXiv.2208.14424} {arXiv:2208.14424 [quant-ph]}.

\end{thebibliography}
\end{document}